\documentclass[%
 reprint,
 amsmath,amssymb,
 aps,
]{revtex4-2}

\usepackage{graphicx}
\usepackage{dcolumn}
\usepackage{bm}
\usepackage[T1]{fontenc}

\usepackage[colorlinks=true,allcolors=blue]{hyperref}
\hypersetup{breaklinks=true}

\begin{document}

\preprint{APS/123-QED}

\title{Strong-Field Nonsequential Double Photoionization\\Using Virtual Detector Theory with Path Summation}

\author{Daniel Younis}
\email{dan.younis@outlook.com}
\author{Joseph H. Eberly}
\affiliation{Center for Coherence and Quantum Optics, and Department of Physics and Astronomy,\\University of Rochester, Rochester, New York 14627, USA}

\date{March 3, 2023}

\begin{abstract}
We present an \emph{ab initio} study of the nonsequential strong-field ionization dynamics of a model two-electron atom with helium character. Single- and double-ionization events are characterized and displayed using detector signals extracted at different points in the two-electron two-dimensional space. The double photoelectron momentum distribution is calculated via coherent path-summation over virtual particle trajectories. Insights into different ionization and electron recollision pathways are gained from detailed virtual-particle tracking and energy-time readouts. This study demonstrates the extension of virtual detector theory to strong-field multi-electron quantum dynamics and highlights the importance of the evolving quantum phase in quasi-classical electron propagation.
\end{abstract}

\maketitle


\emph{Introduction.}~Ionization is the obviously necessary precursor to many strong-field phenomena, including multiphoton \cite{Eberly-book, Delone-book, Smith-book, Fedorov-book}, above-threshold \cite{Agostini-1979, Eberly-1991, Corkum-1993}, and nonsequential multiple ionization \cite{Becker-2005, Becker-2008, Becker-2012}, in addition to high-harmonic x-ray generation \cite{L'Huillier-1991, Lewenstein-1994}, attosecond pulse formation \cite{Krausz-2009, Chini-2014}, and electron recollision spectroscopy \cite{Corkum-2011, Schell-2018}. Many aspects of the single-active-electron ionization process are accessible theoretically on the basis of radiative perturbation theory and the strong-field approximation, or the Keldysh-Faisal-Reiss theory \cite{Keldysh-1964, *Keldysh-1965, Faisal-1973, Reiss-1980} (see \cite{Popruzhenko-2014} for a recent review). In this approach, the scattered electron + field system is treated as a dressed Volkov state \cite{Volkov-1935} and the effect of the atomic potential is included perturbatively.

Nonsequential multiple ionization \cite{Corkum-1993} is characterized by strong inter-electron correlations that can promote cooperative electron exit dynamics \cite{Becker-2005, Becker-2008, Becker-2012}. It is predominantly initiated when a first electron tunnels through the field-suppressed coulombic barrier and, in the next field half-cycle, is accelerated and field-driven back toward the nucleus. Exchange of momentum with residual bound electrons increases the likelihood of collision or tunnel ionization thereafter. The result is an anomalous ionization yield that can greatly exceed what is predicted by the sequential theory \cite{Walker-1994}.

Due to the possibility of such inter-electron correlations, \emph{ab initio} numerical methods are indispensable in laser-atom interaction studies involving field strengths near to or greater than the atomic unit $I\simeq 10^{16}\textrm{ W/cm}^2$. Direct numerical integration of the time-dependent Schr\"{o}dinger equation (TDSE) is the most accurate approach, although it remains intractable beyond 3 wavefunction degrees of freedom \cite{Taylor-1998, Bauer-book} because the number of grid points in the discretely-sampled volume grows exponentially with each added dimension. Attractive alternate schemes include the quantum trajectory method \cite{Wyatt-1999, Wyatt-book}, which represents the wavefunction as a collection of fluid elements obeying the Madelung-Bohm quantum hydrodynamic equations \cite{Madelung-1927, Bohm-1952-1, *Bohm-1952-2}. Another approach is the classical ensemble method \cite{Panfili-2001, Ho-2005-0, Ho-2005-2}, which considers a large collection of particles whose initial conditions and statistical features are determined by the initial wavefunction.

\begin{figure}[b!]
    \centering
    \includegraphics[width=0.38\textwidth]{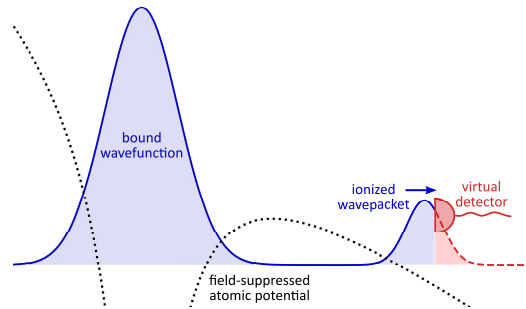}
    \caption{Illustration of the numerical detection process, in which a ``virtual detector'' extracts information from incident wavepackets.}
    \label{fig:schematic}
\end{figure}

More recently, a hybrid quantum-classical approach known as the virtual detector method was initiated by Feuerstein and Thumm \cite{Feuerstein-2003}, which has been extended to include virtual particles \cite{Wang-2013, Wang-2018} and their quantum phase information \cite{Xu-2021}.

In the virtual particle calculation, one introduces an enclosure of purely numerical ``detectors'' around the region where the TDSE is integrated. The exterior of the enclosure is relatively far from the ionization inception, and the result of each detection can be interpreted as the creation of a virtual particle with calculated momentum and phase that carries information about the quantum state, see Fig.~\ref{fig:schematic}. The particle's phased momenta are subsequently treated classically. In this way, the accuracy of the fully quantum-mechanical solution on a grid close to the atomic nucleus can be combined with the efficiency of classical propagation beyond the enclosure, where additional quantum effects are negligible.

This interpretation of the virtual detector method has proven successful in a variety of ways. For example, in analyzing an ionization event independent of a tunneling assumption, it allows one to obtain centrally important features of ionization including those associated with the popular tunneling picture, such as the electron's effective tunneling rate, realistic tunneling entrance and exit positions, and outgoing far-field momentum (see cautions by Ivanov, \emph{et al.}~\cite{Ivanov-2005}, and \cite{Teeny-2016, Tian-2017}). The strong-field recollision scenario relies on determinations of such quantities, which virtual detectors have been able, sometimes uniquely, to provide.

In this Letter, we extend the theory based on virtual detection to two-electron atomic systems. The generalization to three or more interacting electrons readily follows from our reformulation of virtual detector theory. For concreteness, we demonstrate the calculation for a model helium atom and its nonsequential ionization dynamics under strong-field irradiation. We employ the aligned-electron approximation \cite{Javanainen-1988, Haan-2002} wherein the motion of each electron is constrained to the field polarization axis. Thus, our two-dimensional system consists of 2 one-dimensional electrons identified by their positions ($x_1$,\,$x_2$) on two such independent polarization-aligned coordinates.

\emph{Method.}~Along the enclosure net, detectors are densely arranged to intercept the wavefunction $\Psi(\vec{x},t)$ and perform non-destructive numerical detections at their respective positions $\vec{x}_d$ for every calculation time-step $t_d$. The features extracted in the detection are the local phase, probability current, and momentum:
\begin{align}
    \phi &= \arctan(\textrm{Im}\,\Psi/\textrm{Re}\,\Psi)|_{(\vec{x}_d,t_d)} \label{eqn:VD-phase}
    \\
    \vec{j} &= \tfrac{i}{2}(\Psi\nabla\Psi^*-\Psi^*\nabla\Psi)|_{(\vec{x}_d,t_d)} \label{eqn:VD-current}
    \\
    \vec{k} &= \nabla\phi\equiv\vec{j}/\rho\,|_{(\vec{x}_d,t_d)} \label{eqn:VD-momentum}
\end{align}
where $\rho=|\Psi|^2$ is the two-electron probability density. Atomic units are employed unless indicated otherwise. Since the detectors generally lie between numerical grid points, $\Psi(\vec{x},t)$ must be interpolated to each point $\vec{x}_d$.

Equations (\ref{eqn:VD-phase})\,--\,(\ref{eqn:VD-momentum}) initiate a virtual particle at the space-time point of detection $(\vec{x}_d,t_d)$ with initial momentum $\vec{k}$ and a statistical weight $w$ equal to the probability density $\rho$ at birth. Its subsequent motion is governed by Hamilton's classical equations, $\dot{x}_n=\partial H/\partial p_n$, $\dot{p}_n=-\partial H/\partial x_n$ for $n=1,2$. Thus, the representation of outward-bound wavepackets is converted from a quantum wave to a classical particle density description, and likewise the Hamiltonian changes from a quantum operator to a classical function as the interpretation switches. The evolution of one virtual particle in ($x_1$,\,$x_2$)-space represents the dynamical behavior of two classical electrons, and its trajectory signifies a possible two-electron ionization pathway from the ensemble of cases.

Under irradiation by high-intensity, low-frequency, and/or long-duration pulses, the spatial domain may not be able to accommodate the Schr\"{o}dinger wavefunction far from the nucleus. Thus, at the domain boundary one typically employs a masking function or negative complex potential \cite{Ge-1998} to numerically absorb ionized wavefunction components. In contrast, the virtual-particle description of the wavefunction is not restricted to a grid. Moreover, virtual particles do not interact (but the two electrons they represent interact pairwise), so their time-evolution also offers the advantage of computation in parallel.

\begin{figure}[t!]
    \centering
    \includegraphics[width=0.35\textwidth]{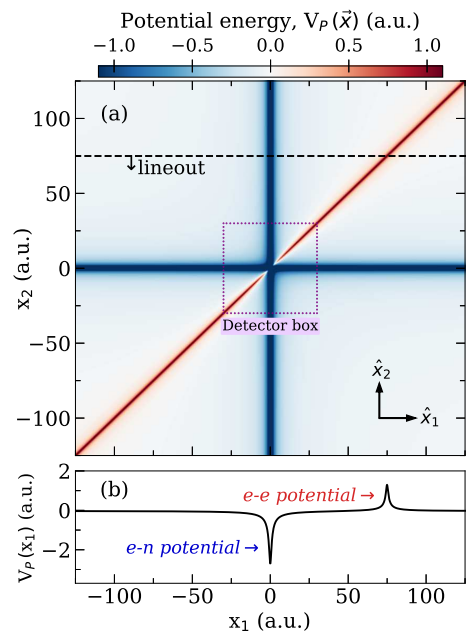}
    \caption{(Color online). (a) Potential energy map of the model two-electron atom with the virtual detector box overlaid. One-dimensional interaction potentials are obtained from lineouts parallel to the coordinate axes $\hat{x}_1$/$\hat{x}_2$, as in panel (b).}
    \label{fig:potential-energy-map}
\end{figure}

The Hamiltonian
\begin{align}
    H(\vec{x},\vec{p},t) = \sum_{n=1}^{2}\bigg[&\frac{1}{2}\,p^2_n - Z_0 V(x_n)\bigg] \label{eqn:Hamiltonian}
    \\
    & + V(x_1-x_2) + (x_1+x_2) E(t) \nonumber
\end{align}
consists of electron-nuclear ($e$-$n$) and electron-electron ($e$-$e$) screened coulombic potentials of the form $V(x)=1/\sqrt{x^2+\sigma^2}$, in addition to a length-gauge field interaction term under the dipole approximation \cite{Javanainen-1988, Haan-2002}. Here, $Z_0$ is the nuclear charge, $E(t)$ is the laser electric field, and $\sigma$ is a screening parameter. The non-zero value for $\sigma$ controls the $1/x$ coulombic singularity in $V(x)$, and it also determines the model atomic spectrum and ionization energy.

\begin{figure*}[t!]
    \centering
    \includegraphics[width=\textwidth]{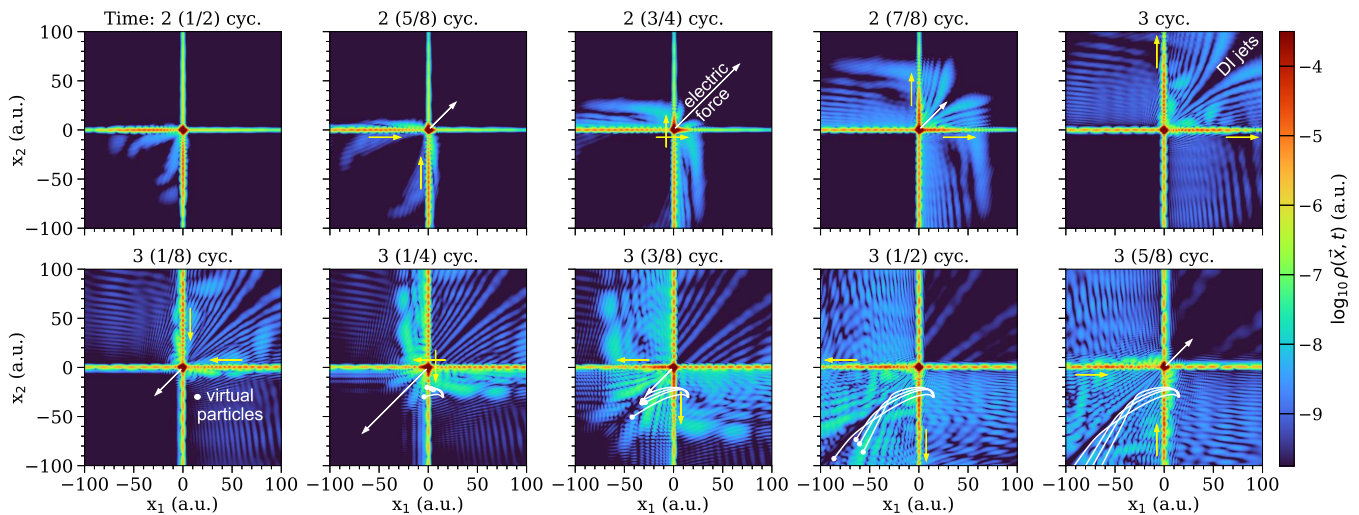}
    \caption{(Color online). Snapshots of the two-electron probability density $\log_{10}\rho(\vec{x},t)$. White arrow: instantaneous laser electric force vector, $-(\hat{x}_1+\hat{x}_2)E(t)$. Yellow arrows: direction of probability current flow on-axis. The trajectories of four virtual particles initiated at 3 (1/8) cyc.~are shown, which subsequently follow the stream of probability current.}
    \label{fig:2e-density-evolution}
\end{figure*}

\emph{Calculation parameters.}~We set $Z_0=+2$ and $\sigma=0.74\textrm{ a.u.}$~for a ground-state energy of $-2.902\textrm{ a.u.}$~($\approx-79\textrm{ eV}$) corresponding closely to that of helium. Figure \ref{fig:potential-energy-map} illustrates the spatial dependence of the total atomic potential energy and includes the detector box. It has a side-length of $60\textrm{ a.u.}$~and consists of 500 detectors distributed uniformly along its perimeter. The separation between neighboring detectors is chosen to be finer than the spatial resolution of the discrete wavefunction, making the enclosure effectively complete in the sense that probability current density does not pass the detector box unregistered.

The wavelength of the field is $780\textrm{ nm}$ and its peak intensity is $1/2\textrm{ PW/cm}^2$. These parameters correspond to the so-called nonsequential double ionization ``knee'' process in helium, where the $e$-$e$ correlation strength is enhanced \cite{Walker-1994, Liu-1999}. We take the temporal profile of the laser pulse amplitude to be trapezoidal with a 6-cycle plateau period and a 2-cycle linear turn-on and turn-off, equaling a total pulse duration of $26\textrm{ fs}$.

\emph{Results: Wavefunction dynamics \& detector signals.}~In this coordinate representation, the two-electron wavefunction is attracted to the $x_1$ and $x_2$ axes by the nuclear potential, and it is repelled away from the $x_1=x_2$ diagonal due to inter-electron repulsion (see Fig.~\ref{fig:potential-energy-map}). With every half-cycle, the field tilts the total potential experienced by both electrons, energetically raising or lowering it toward either the positive or negative side of their axes. The four quadrants of the position-space are readily understood: population in the $x_1 x_2 > 0$ regions signifies a non-zero probability of detecting both electrons on the same side of the nucleus, and conversely for population in the $x_1 x_2 < 0$ regions. Additionally, near-axis population far from the origin indicates that one electron is bound while the other is well ionized.

A time-sequence of the 2$e$ probability density $\rho(\vec{x},t)$ is shown in Fig.~\ref{fig:2e-density-evolution}. Here, one observes the formation of double-ionization (DI) jets every half-cycle, corresponding to the ejection of both electrons in the same direction $x_1 x_2 > 0$. Alternatively, probability density develops in the $x_1 x_2 < 0$ quadrants primarily due to sequential field-ionization, i.e., the electrons tunneling in opposite directions during opposite field half-cycles. The subsequent dynamics can be understood from the trajectory evolution of virtual particles. Consider the four particles born at $t_d=\textrm{3 (1/8) cyc.}$~around $\vec{x}_d=(15,-30)\textrm{ a.u.}$ as shown in Fig.~\ref{fig:2e-density-evolution}, all of which represent sequentially-ionized electron pairs. As the field reverses direction, the trajectories illustrate how the first electron is driven back toward the origin after which it scatters off the nuclear potential and contributes to the double-ionizing jet population. This reveals that double-ionization jets are not formed solely because of both electrons tunneling out in the same laser half-cycle.

\begin{figure}[b!]
    \centering
    \includegraphics[width=0.475\textwidth]{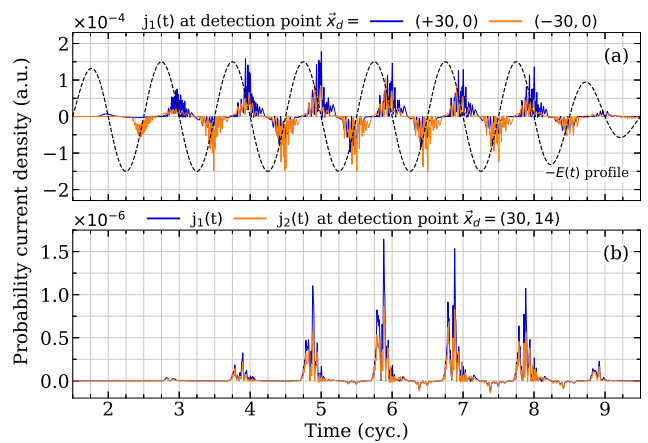}
    \caption{(a) Time series of the $\hat{x}_1$-directed probability current density measured 30 a.u.~to the right/left (in blue/orange) of the atomic core. Dashed black line: laser electric force profile. (b) Time series of the $\hat{x}_1$/$\hat{x}_2$ components of $\vec{j}(t)$ (in blue/orange) measured at the detection point $(30,14)\textrm{ a.u.}$}
    \label{fig:current-time-series}
\end{figure}

The virtual detector signals also provide insight into the ionization process, particularly regarding the timing of events. In Fig.~\ref{fig:current-time-series}(a), the readouts of axial probability current $j_1(t)=\vec{j}(t)\cdot\hat{x}_1$ from the left- and right-most detectors at $\vec{x}_d=(\pm30,0)$ a.u.~are provided. With each laser half-cycle, a probability current signal of comparable duration is registered which lags the field crest by $\sim$\,1/4th of a cycle. These axial, inward-directed current signals are of particular relevance to the recollision scenario of strong-field ionization, for they are associated with the virtual particles of bound-recolliding electron pairs (their dynamical behavior will be analyzed in what follows). In Fig.~\ref{fig:current-time-series}(b), the two components of $\vec{j}(t)$ are provided for the detector at $\vec{x}_d=(30,14)\textrm{ a.u.}$, which according to Fig.~\ref{fig:2e-density-evolution} is in the path of a double-ionization jet. In this case, the probability current signal lags the electric field crest by $\sim$\,1/8th of a cycle.

The $j_n(t)$ signals provided in Figs.~\ref{fig:current-time-series}(a) and \ref{fig:current-time-series}(b) represent single- and double-ionization events, respectively, due to their associated detection points in space. Comparing the signal amplitudes, it is seen that the current density for single ionization is approximately two orders of magnitude stronger than that for double ionization. Lastly, the sub-cycle oscillations in $j_n(t)$ are due to spatiotemporal wavefunction interference, and the signals are modulated by the laser pulse profile in addition to the depletion of bound population over time.

\emph{Results: Photoelectron momentum distribution.}~The 2-$e$ probability density exhibits a complex spatial interference structure arising from field-driven collisions between different wavepacket components. The virtual detector method captures this information by associating to each particle an initial phase, given by the local wavefunction phase at birth, and tracking its evolution \cite{Xu-2021}. This brings the quantum-classical correspondence between the wavefunction and virtual-particle descriptions closer.

The phase calculation is based on the observation that the quantum wave of a virtual particle may be approximated by a Volkov state \cite{Volkov-1935, Popruzhenko-2014}: $\Psi_\textsc{v}(\vec{x},t)=(2\pi)^{-1}\exp(i\vec{k}\cdot\vec{x})\exp[-iS(t)]$ where $\vec{k}$ is the wavevector and $S(t)=\int dt\,L$ is the action integral of Lagrangian $L$. In accordance with $\Psi_\textsc{v}(\vec{x},t)$, the evolving virtual-particle phase is $\phi(t)=\phi_0-\vec{k}(t)\cdot\vec{x}(t)+S(t)$ where $\phi_0$ is the initial phase and $\vec{k}(t)$ is now the instantaneous momentum.

The photoelectron momentum distribution (PMD) is calculated by binning the virtual particle weights with their path-integrated phase terms: $W(\vec{p})=|\sum_{n}\sqrt{w_n}\,\exp(i\phi_n)|^2$ where $\vec{p}_n\in[\,\vec{p},\vec{p}+\delta\vec{p}\,)$, $\delta\vec{p}$ is the vector of bin widths. Virtual particles for which either electron is still bound to the nucleus are omitted from the summation to produce the distribution of doubly-ionized electron pairs. From the final position-space wavefunction $\Psi(\vec{x})$, the photoelectron momentum distribution can also be calculated via Fourier transformation after applying a suitable masking function $\mathcal{M}(\vec{x})$ to filter the bound population: $\Phi(\vec{p})=\frac{1}{2\pi}\int d\vec{x}\,e^{-i\vec{p}\cdot\vec{x}} (\mathcal{M}\Psi)(\vec{x})$. In this case, $\mathcal{M}(\vec{x})$ is a cross-shaped gaussian filter that smoothly attenuates the bound population and all singly-ionized wavepackets. Further, a cross-shaped momentum-space filter was applied to $\Phi(\vec{p})$ to remove population for which either photoelectron momentum is low.

\begin{figure}[t!]
    \centering
    \includegraphics[width=0.475\textwidth]{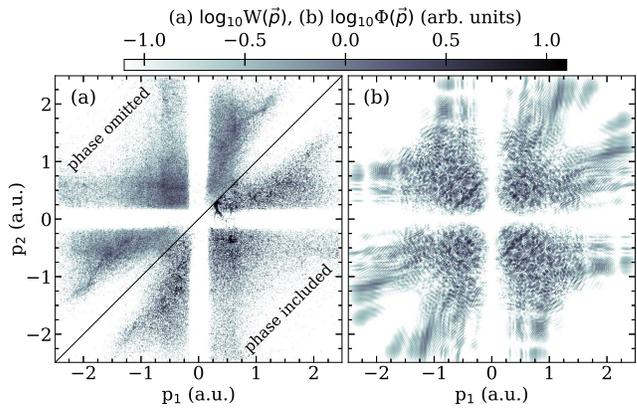}
    \caption{The double-photoelectron momentum distribution calculated using (a) $54\textrm{ million}$ virtual electron pairs in $W(\vec{p})$, and (b) the free-space wavefunction $\Phi(\vec{p})$. The upper half of panel (a) illustrates the effect of neglecting the virtual-particle phase.}
    \label{fig:pmd}
\end{figure}

In Fig.~\ref{fig:pmd}, the double-photoelectron momentum distributions, calculated using $W(\vec{p})$ and $\Phi(\vec{p})$, are provided. The phase-included half of $W(\vec{p})$ is sharper than its phase-omitted counterpart in panel (a), and it reveals more of the speckled interference pattern exhibited also by $\Phi(\vec{p})$. The virtual-particle distribution is also in excellent qualitative agreement with that obtained by the classical ensemble method in a similar intensity/wavelength regime (cf.~Fig.~3 in Ref.~\cite{Ho-2005-1}). However, the resolution is much finer in this case due to the greater number of electron pairs that comprise the distribution, which is on the order of $10^7$, versus the $10^5$ members used in Ref.~\cite{Ho-2005-1}.

\begin{figure*}[t!]
    \centering
    \includegraphics[width=\textwidth]{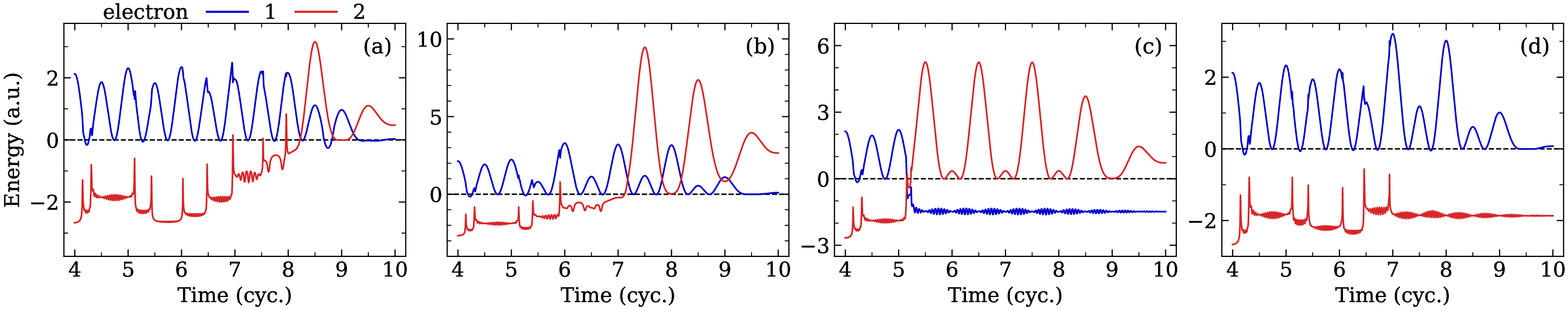}
    \caption{(Color online). Energy-time evolution of sample virtual electron pairs. Initially, electron 1 (in blue) is ionized, while electron 2 (in red) is bound. Inter-electron collisions appear as sharp cusps, most prominently seen in the bound electron curves. The four panels show (a) nonsequential double ionization, (b) recollision-excitation with subsequent ionization, (c) bound-ionized electron swapping, and (d) failure to double-ionize after multiple collisions.}
    \label{fig:particles-energy-time}
\end{figure*}

In calculations involving single-active-electron atoms, interference rings appear in the PMDs whose radii are integer multiples of $\sqrt{2\omega}$ \cite{Xu-2021}. Thus, they correspond energetically to local maxima in the above-threshold ionization spectrum. In the two-electron case, the $e$-$e$ interaction further complicates the energy transfer mechanisms in the atom-field system, and the ripple-like pattern in Fig.~\ref{fig:pmd} does not have a direct interpretation. Lastly, the maximum cutoffs in the PMDs correspond approximately to $E_0/\omega\approx 2\textrm{ a.u.}$, which is the momentum amplitude of a classical electron oscillating in a plane electromagnetic wave.

\emph{Results: Virtual electron trajectories.}~The virtual particle dynamical variables can be tracked in time, providing a classical view into the ionization and $e$-$e$ interaction processes that is similar to the classical ensemble method and is reminiscent of alternate formulations of Schr\"{o}dinger theory such as Bohmian mechanics \cite{Bohm-1952-1, *Bohm-1952-2} and the Feynman path-integral approach \cite{Feynman-book}.

In this calculation, there are $\sim$\,50 million virtual particles, equal to the product of the total number of detectors and discrete time-steps used. Figure \ref{fig:particles-energy-time} shows the energy-time evolution of a few representative virtual electron pairs which undergo multiple collision events. They were each initialized at $(x_1,x_2)=(-30,0)\textrm{ a.u.}$~a few time-steps apart, beginning around the 4th field cycle. According to Fig.~\ref{fig:current-time-series}(a) (orange curve), the probability current at this position and time is flowing in the positive direction toward the nucleus, which signals an upcoming bound-free $e$-$e$ collision.

In all cases, the first electron is ionized while the second electron is still bound. In Table \ref{tab:electron-ep}, the range of initial energies and momenta of each virtual electron pair is provided to convey their proximity. However, their subsequent dynamical behavior is significantly different, as evidenced in Fig.~\ref{fig:particles-energy-time}, revealing the high degree of sensitivity of the interaction on the initial conditions which are derived from the wavefunction. For instance, Fig.~\ref{fig:particles-energy-time}(a) illustrates the process of nonsequential double ionization (NSDI) \cite{Walker-1994, Ho-2005-0, Liu-1999} in which a series of energetically favorable collisions occurring approximately every half-cycle causes the bound electron to transition into the continuum. Figure \ref{fig:particles-energy-time}(b) shows the related process of recollision-excitation with subsequent ionization (RESI) \cite{Hao-2022} in which, following a collision event, the bound electron occupies an excited intermediate state (in this case, between $t\approx 6-7\textrm{ cyc.}$) from which it is later field-ionized.

\begin{table}[t!]
\caption{\label{tab:electron-ep}The range of initial energies and momenta of the virtual electron pairs shown in Fig.~\ref{fig:particles-energy-time}.}
\begin{ruledtabular}
\renewcommand{\arraystretch}{1.15}
\begin{tabular}{l|cc}
\textit{Electron no.} & \textit{1 (in blue)} & \textit{2 (in red)} \\
\hline
\textit{Energy} & $2.13\pm10^{-2}\textrm{ a.u.}$ & $-2.67\pm10^{-2}\textrm{ a.u.}$ \\
\textit{Momentum} & $2.07\pm10^{-2}\textrm{ a.u.}$ & ($-4.3 \pm 0.2)\times10^{-2}\textrm{ a.u.}$ \\
\end{tabular}
\end{ruledtabular}
\end{table}

Figure \ref{fig:particles-energy-time}(c) illustrates a situation in which the bound and ionized electrons swap following interaction, resulting in a singly-ionized atomic state. In this case, the first electron is recaptured by the nucleus after thrice-colliding with the second electron, which emerges in the continuum with more than double the initial energy of the former. Lastly, Fig.~\ref{fig:particles-energy-time}(d) shows multiple $e$-$e$ collision events that ultimately fail to liberate the bound electron. This suggests that double ionization is more sensitive to the timing of energy transfers, between the electrons themselves and the electrons with the field, than it is to the overall number of collision events. Evidently, the diverse range of correlated inter-electron behavior can be interpreted in a direct way using the virtual detector method.

\emph{Summary.}~In conclusion, we have demonstrated how the virtual detector method can be applied to probe the evolution of a two-electron atom and its nonsequential ionization dynamics arising from strong-field irradiation. The detector signals and virtual particle dynamical variables provide valuable insights into the behaviors leading up to single- and double-ionization events. Furthermore, the virtual particle momentum distribution including path-integrated phase information agrees qualitatively with the full quantum-mechanical solution based on numerical integration of the time-dependent Schr\"{o}dinger equation. The possibility of applying the virtual detector method to elucidate other ionization-related processes may be explored in the near future.

\begin{acknowledgments}
The work reported here was supported by the grant DE-FG02-05ER15713 funded by the U.S.~Department of Energy, Office of Science. Calculations were performed on the BlueHive supercomputing cluster at the University of Rochester.
\end{acknowledgments}
\clearpage

\providecommand{\noopsort}[1]{}\providecommand{\singleletter}[1]{#1}%

\end{document}